\documentclass[english,letter paper,12pt,reqno]{article}
\usepackage{EXECUTION}
\usepackage{hyperref}

\usepackage{color, colortbl}
\usepackage{caption}
\definecolor{Gray}{gray}{0.9}

\usepackage{caption}
\usepackage{subcaption}

\begin{document}

\title{\textbf{Transformers for limit order books}}
\author{James Wallbridge\footnote{Correspondence to james.wallbridge@gmail.com}\\
}
\date\today{\normalsize}
\maketitle

\begin{center}
\textbf{Abstract}
\end{center}

We introduce a new deep learning architecture for predicting price movements from limit order books.  This architecture uses a causal convolutional network for feature extraction in combination with masked self-attention to update features based on relevant contextual information.  This architecture is shown to significantly outperform existing architectures such as those using convolutional networks (CNN) and Long-Short Term Memory (LSTM) establishing a new state-of-the-art benchmark for the FI-2010 dataset. \\

\tableofcontents

\section{Introduction}

Understanding high-frequency market micro-structure in time-series data such as limit order books (LOB) is complicated by a large number of factors including high-dimensionality, trends based on supply and demand, order creation and deletion around price jumps and the overwhelming relative percentage of order cancellations.  It makes sense in this inherently noisy environment to take an agnostic approach to the underlying mechanisms inducing this behavior and construct a network which learns to uncover the relevant features from raw data.  This removes the bias contained in models using hand-crafted features and other market assumptions such as those in autoregressive models VAR \cite{Zi} and ARIMA \cite{Ar}.

Arguably the most successful architecture used to extract features is the convolutional neural network \cite{Le} which makes use of translation equivariance, present in many domains including time-series applications.  For time-series however, further inductive biases prove to be beneficial.  Convolutional neural networks with a causal temporal bias were introduced in \cite{Oo} to encode long-range temporal dependencies in raw audio signals.  Here convolutions are replaced by dilated causal convolutions controlled by a dilation rate.  The dilation rate is the number of input values skipped by the filter, thereby allowing the network to act with a larger receptive field.  In this work, features from our architecture will come from the output of multiple such dilated causal convolutional layers connected in series.

Once we have a collection of features, we would like to do computations with these learned representations to enable context dependent updates.  Historically, attention networks were introduced in \cite{Bah} to improve existing long-short term memory (LSTM) \cite{Ho,Gr} models for neural machine translation by implementing a ``soft search" over neighboring words enabling the system to focus only on words relevant to the generation of the next target word.  This early work combined attention with RNNs.  Shortly after, CNNs were combined with attention in \cite{Xu} and \cite{Ch} for image captioning and question-answering tasks respectively.  

In \cite{Va}, self-attention was introduced as a stand alone replacement for LSTMs on a wide range of natural language processing tasks leading to state-of-the-art results \cite{De,Ra} which included masked word prediction.  Introducing self-attention can be thought of as incorporating an inductive biases into the learning architecture to exploit relational structure in the task environment.  This amounts to learning over a graph neural network \cite{Sc,Bat} where nodes are entities given by the learned features which are then updated through a message passing procedure along edges.  Results in various applications show that self-attention can better capture long range dependencies in comparison to LSTMs \cite{Da}.  

More precisely, \cite{Va} introduced the transformer architecture which consists of an encoder and decoder for language translation.  Both the encoder and decoder contain the repetition of modules which we refer to as \textit{transformer blocks}.  Each transformer block consists of a multi-head self-attention layer followed by normalization, feedforward and residual connections.  This is described in detail in Section~\ref{section:architecture}.  

Combining transformer blocks with convolutional layers for feature extraction is a powerful combination for various tasks.  In particular, for complex reasoning tasks in various strategic game environments, the addition of these transformer modules significantly enhanced performance and sample efficiency compared with existing non-relational baselines \cite{Za,Vi,Cl}.  In this work we combine the causal convolutional architecture of \cite{Oo} with multiple transformer blocks.  Moreover, our transformer blocks contain masked multi-head self-attention layers.  By applying a mask to our self-attention functions, we ensure that the ordering of events in our time-series is never violated at each step, ie. entities can only attend to entities in its causal past.  

We train and test our model on the publicly available FI-2010 data-set\footnote{The ``MNIST" for limit order books.} which is a LOB of five instruments from the Nasdaq Nordic stock market for a ten day period \cite{Nt}.  We show that our algorithm outperforms other common and previously state-of-the-art architectures using standard model validation techniques.  

In summary, inspired by the wavenet architecture of \cite{Oo} where dilated causal convolutions were used to encode long-range temporal dependencies, we use these causal convolutions to build a feature map for our transformer blocks to act on.  We refer to our specific architecture as TransLOB.  It is a composition of differentiable functions that process and integrate both local and global information from the LOB in a dynamic relational way whilst respecting the causal structure.  

There are a number of advantages to our architecture outside of the significant increases in performance.  Firstly, in spite of the $O(N^2)$ complexity of the self-attention component, our architecture is substantially more sample efficient than existing LSTM architectures for this task.  Secondly, the ability to analyse attention distributions provides a clearer picture of internal computations within the model compared with these other methods leading to better interpretability.

\section*{Related work}

There is now a substantial literature applying deep neural networks to time-series applications, and in particular, limit order books (LOB).  Convolutional neural networks (CNN) have been explored in LOB applications in \cite{Do,Ts}.  To capture long-range dependencies in temporal behavior, CNNs have been combined with recurrent neural networks (RNN) (typically long-short term memory (LSTM)) which improve on earlier results \cite{Ts2,Zh}.  Some modifications to the standard convolutional layer have been used in attempts to infer local interactions over different time horizons.  For example, \cite{Zh} uses an inception module \cite{Sz} after the standard convolutional layers for this inference followed by an LSTM to encode relational dynamics.  Stand-alone RNNs have been used extensively in market prediction \cite{Di,Fi,Bao} and have been shown to outperform models based on standard multi-layer perceptrons, random forests and SVMs \cite{Ts3}.  

For time-series applications, recent work \cite{Tr,Qi} uses attention and \cite{La,Ma,Sh} in combination with CNNs.  However, there are relatively few references which combine CNNs with transformers to analyse time-series data.  We mention \cite{So} which uses a CNN plus multi-head self-attention to analyse clinical time-series behaviour and \cite{Li} which became aware to us during the final write-up of this paper which uses a similar architecture to our own and applied to univariate synthetic and energy sector datasets.  As far as we are aware, ours is the first work applying this class of architectures to the multivariate financial domain, with the various subtleties arising in this particular application.

\section{Experiments}

A limit order book (LOB) at time $t$ is the set of all active orders in a market at time $t$.  These orders consist of two sides; the bid-side and the ask-side.  The bid-side consists of buy orders and the ask-side consists of sell orders both containing price and volume for each order.  Our experiments will use the LOB from the publicly available FI-2010 dataset\footnote{The dataset is available at https://etsin.fairdata.fi/dataset/73eb48d7-4dbc-4a10-a52a-da745b47a649}.  A general introduction to LOBs can be found in \cite{Go}.

Let $\{p_a^{i}(t),v_a^{i}(t)\}$ denote the price (resp. volume) of sell orders at time $t$ at level $i$ in the LOB.  Likewise, let $\{p_b^{i}(t),v_b^{i}(t)\}$ denote the price (resp. volume) of buy orders at time $t$ at level $i$ in the LOB.  The bid price $p_b^{1}(t)$ at time $t$ is the highest stated price among active buy orders at time $t$.  The ask price $p_a^{1}(t)$ at time $t$ is the lowest stated price among active sell orders at time $t$.  A buy order is executed if $p_b^{1}(t)>p_a^{1}(t)$ for the entire volume of the order.   Similarly, a sell order is executed if $p_a^{1}(t)<p_b^{1}(t)$ for the entire volume of the order.  

The FI-2010 dataset is made up of 10 days of 5 stocks from the Helsinki Stock Exchange, operated by Nasdaq Nordic, consisting of 10 orders on each side of the LOB.  Event types can be executions, order submissions, and order cancellations and are non-uniform in time.  We restrict to normal trading hours (no auction).  The general structure of the LOB is contained in Table~\ref{table:lobtable}.  

\begin{table}[!htb]\tiny
\renewcommand{\arraystretch}{1.7}
   \begin{minipage}{.3\linewidth}
      \centering
    \begin{tabular}{ | c |}
    \hline 
    $(p^{10}_a(t),v^{10}_a(t))$ \\ \hline
        \vdots \\ \hline
    $(p^1_a(t),v^1_a(t))$ \\ \hline \rowcolor{Gray}
    \\ \hline
    $(p^1_b(t),v^1_b(t))$ \\ \hline
    \vdots\\ \hline    
    $(p^{10}_b(t),v^{10}_b(t))$ \\
    \hline
    \end{tabular}
\caption*{\footnotesize Event $t$}
\end{minipage}%
    \begin{minipage}{.2\linewidth}
      \centering
  \begin{tabular}{ | c |}
    \hline 
    $(p^{10}_a(t+1),v^{10}_a(t+1))$ \\ \hline
        \vdots \\ \hline
    $(p^1_a(t+1),v^1_a(t+1))$ \\ \hline \rowcolor{Gray}
    \\ \hline
    $(p^1_b(t+1),v^1_b(t+1))$ \\ \hline
    \vdots\\ \hline    
    $(p^{10}_b(t+1),v^{10}_b(t+1))$ \\
    \hline
    \end{tabular}
\caption*{\footnotesize Event $t+1$}    
\end{minipage}%
 \begin{minipage}{.2\linewidth}
      \centering
  \begin{tabular}{ c }  
    \\ 
\ldots\ldots\ldots\ldots\ldots\ldots\ldots\ldots \\ 
     \\
    \\ 
     \\ 
    \\   
     \\
    \end{tabular}  
\end{minipage}%
    \begin{minipage}{.2\linewidth}
      \centering
    \begin{tabular}{ | c |}
    \hline 
    $(p^{10}_a(t+10),v^{10}_a(t+10))$ \\ \hline
        \vdots \\ \hline
    $(p^1_a(t+10),v^1_a(t+10))$ \\ \hline \rowcolor{Gray}
    \\ \hline
    $(p^1_b(t+10),v^1_b(t+10))$ \\ \hline
    \vdots\\ \hline    
    $(p^{10}_b(t+10),v^{10}_b(t+10))$ \\
    \hline
    \end{tabular}
        \caption*{\footnotesize Event $t+10$}    
\end{minipage} 
 \caption{Structure of the limit order book.}
\label{table:lobtable}
 \end{table}

The data is split into 7 days of training data and 3 days of test data.  Preprocessing consists of normalizing the data $x$ according to the \textit{$z$-score} 
\[   \bar{x}_t = \frac{x_t-\bar{y}}{\sigma_{\overline{y}}}  \]
where $\overline{y}$ (resp. $\sigma_{\overline{y}}$) is the mean (resp. standard deviation) of the previous days data.  Since the aim of this work is to extract the most amount of possible latent information contained in the LOB, we do not include any of the hand-crafted features contained in the FI-2010 dataset.  For a detailed description of this dataset we refer the reader to \cite{Nt}.  

We aim to predict future movements from the (virtual) mid-price.  Price direction of the data is calculated using the following smoothed version of the mid-price.  This amounts to adjusting for the average volatility of each instrument.  The virtual mid-price is the mean
\[  p(t) = \frac{p_a^{1}(t) + p_b^{1}(t)}{2}  \]
between the bid-price and the ask-price.  The mean of the next $k$ mid-prices is then
\[
m^{+}_k(t) = \frac{1}{k}\sum_{n=0}^k p({t+n}).
\]

The direction of price movement for the FI-2010 dataset is calculated using the percentage change of the virtual mid-price according to 
\[  r_k(t)= \frac{m^+_k(t) - p(t)}{p(t)} .   \]
There exist other more sophisticated methods to determine the direction of price movement at a given time.  However, for fair comparison to other work, we utilize this definition and leave other methods for future work.  The direction is up $(+1)$ if $r_k(t) > \alpha$, down $(-1)$ if $r_k(t) < -\alpha$ and neutral $(0)$ otherwise, according to a chosen threshold $\alpha$.  For the FI-2010 dataset, this has been set to $\alpha=0.002$.  

We consider the following four test cases $k\in\{10, 20, 50, 100\}$ for the denoising horizon window.  The 100 most recent events are used as input to our model.

\section{Architecture}\label{section:architecture}

In this section we give a detailed account of our architecture.  The main two components are a convolutional module and a transformer module.  They contain multiple iterations of dilated causal convolutional layers and transformer blocks respectively.  A transformer block consists of a specific combination of multi-head self-attention, residual connections, layer normalization and feedforward layers.  We took seriously the causal nature of the problem by implementing both causality in the convolutional module and causality in the transformer module through masked self-attention to accurately capture temporal information in the LOB.  Our resulting architecture will be referred to as TransLOB.   

Since each order consists of a price and volume, a state $x_t=\{p_a^i(t),v_a^i(t),p_b^i(t),v_b^i(t)\}_{i=1}^{10}$ at time $t$ is a vector $x_t\in\bb{R}^{40}$.  Events are irregularly spaced in time and the 100 most recent events are used as input resulting in a normalized vector $X\in\bb{R}^{100\times 40}$.  

We apply five one-dimensional convolutional layers to the input $X$, regarded as a tensor of shape $[100,40]$ (ie. an element of $\bb{R}^{100}\otimes\bb{R}^{40}$).  All layers are dilated causal convolutional layers with $14$ features, kernel size $2$ and dilation rates $1, 2, 4, 8$ and $16$ respectively.  This means the filter is applied over a window larger than its length by skipping input values with a step given by the dilation rate with each layer respecting the causal order.  The first layer with dilation rate $1$ corresponds to standard convolution.  All activation functions are $\textup{ReLU}$.  

The full size of the channel filter is used to allow the weights in the filter to infer the relative importance of each level on each side of the mid-price.  It is expected that higher weights will be allocated to shallower levels in the LOB since those levels are most indicative of future activity.  The output of the convolutional module is a tensor of shape $[100,14]$.  

This output then goes through layer normalization \cite{Ba} to stabilize dynamics before each feature vector is concatenated with a one-dimensional temporal encoding resulting in a tensor $X$ of shape $[100,15]$.  We will refer to $N=100$ as the number of \textit{entities} and $d=15$ as the model dimension.  We denote these entities by $e_i$, $1\leq i\leq N$, where $e_i\in E=\bb{R}^{d}$.  These entities are then updated through learning in a number of steps.  

First we introduce an inner product space $H=\bb{R}^d$ with dot product pairing $\<h,h'\>=h\cdot h'$.  We employ a multi-head version of self-attention with $C$ channels.  Therefore, we choose a decomposition $H=H_1\oplus\ldots\oplus H_C$ and apply a linear transformation
\[  T=\bigoplus_{a=1}^C T_a:E\ra\bigoplus_{a=1}^C H_a^{\oplus 3}  \]
with $H_a$ each of dimension $d/C$.   The vectors $(q_{i,(a)},k_{i,(a)},v_{i,(a)})=T_a(e_i)$ are referred to as \textit{query}, \textit{key} and \textit{value} vectors respectively.  We arrange these vectors into matrices $Q_a$, $K_a$ and $V_a$ respectively with $N$-rows and $d$-columns.  In other words, $Q_a=XW_a^Q$, $K_a=XW_a^K$ and  $V_a=XW_a^V$ for weight matrices $W^Q_a$, $W^K_a$ and $W^V_a$ which are vectors in $\bb{R}^{d\times d/C}$.

Next we apply the masked scaled dot-product self-attention function
\[    \textup{head}_a=V_a' = \textup{Softmax}\left(\textup{Mask}\left(\frac{Q_a K_a^T}{\sqrt{d}}\right)\right)V_a   \]
resulting in a matrix of refined value vectors for each entity.  Here $\textup{Mask}$ substitutes infinitesimal values to entries in the upper right triangle of the applied matrix which forces queries to only pay attention to keys in its causal history via the softmax function.  The heads are then concatenated and a final learnt linear transformation is given leading to the multi-head self-attention operation
\[   \textup{MultiHead}(X)=\left(\bigoplus_{a=1}^C\textup{head}_a\right)W^O   \]
where $W^O\in\bb{R}^{d\times d}$.

We next add a residual connection and apply layer normalization resulting in
\[   Z = \textup{LayerNorm}(\textup{MultiHead}(X) + X ).  \]
This is followed by a feedforward network $\textup{MLP}$ consisting of a ReLU activation between two affine transformations applied identically to each position, ie. individually to each row of $Z$.  The inner layer is of dimension $4\times d=60$. Finally, a further residual connection and final layer normalization is applied to arrive at our updated matrix of entities
\[  \textup{TransformerBlock}(X) =  \textup{LayerNorm}(\textup{MLP}(Z) + Z).  \]
The output of the transformer block is the same shape $[N,d]$ as the input.   Our updated entities are $e'_i\in\bb{R}^{15}$, $1\leq i\leq N$.

\begin{figure}[!htb]
\begin{center}
\includegraphics[scale=0.7]{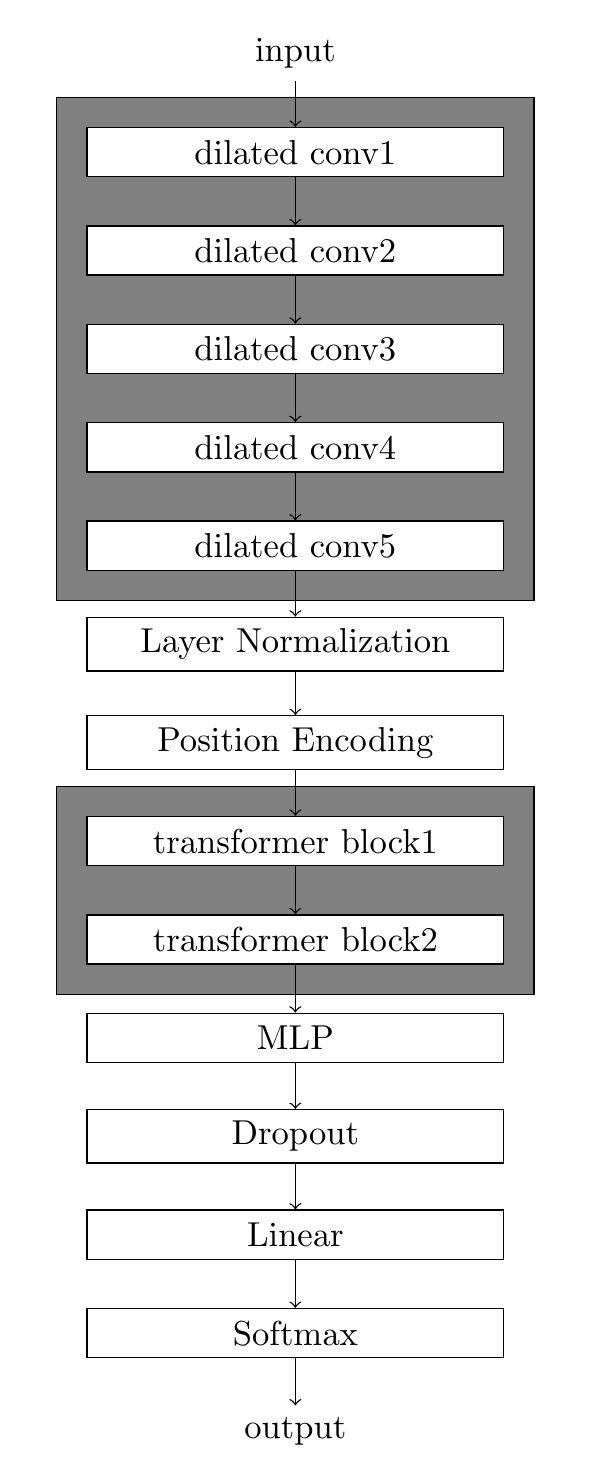}  
\end{center}
\caption{Architecture schematic with enclosed convolutional and transformer modules.}
\label{figure:schema}
\end{figure}

After multiple iterations of the transformer block, the output is then flattened and passed through a feedforward layer of dimension $64$ with ReLU activation and L2 regularization.  Finally, we apply dropout followed by a softmax layer to obtain the final output probabilities.  A schematic of the TransLOB architecture is given in Figure~\ref{figure:schema}. 

For the FI-2010 dataset, we employ two transformer blocks with three heads and with the weights shared between iterations of the transformer block.  The hyperparameters are contained in Table~\ref{table:hyper}.  No dropout was used inside the transformer blocks. 

\begin{table}[!ht]
\footnotesize
    \begin{center}
    \begin{tabular}
    {ll}
    \toprule
      \textbf{Hyperparameter}  & \textbf{Value} \\ 
    \midrule
    Batch size & 32\\
    Adam $\beta_1$ & 0.9\\
    Adam $\beta_2$ & 0.999\\
    Learning rate & $1 \times 10^{-4}$\\
    Number of heads  & 3 \\
    Number of blocks & 2\\
    MLP activations & ReLU \\    
    Dropout rate & 0.1\\
    \bottomrule
    \end{tabular}
    \end{center}
    \caption{Hyperparameters for the FI-2010 experiments.}
    \label{table:hyper}
\end{table}

\section{Results}

Here we record our experimental results for the FI-2010 dataset.  The first 7 days were used to train the model and the last 3 days were used as test data.  Training was done with mini-batches of size 32.  Our metrics include accuracy, precision, recall and F1.  All training was done using one K80 GPU on google colab.  

To be consistent with earlier works using the same dataset, we train and test our model on the horizons $k=\{10,20,50,100\}$.  All models were trained for 150 epochs, although convergence was achieved significantly earlier.  See Figure~\ref{figure:acc_translob} of Appendix~\ref{appendix:baseline} for an example.

The following models were used as comparison.  An LSTM was utilized and compared to a support vector machine (SVM) and multi-layer perceptron (MLP) in \cite{Ts3} with favourable results.  Results using a stand-alone CNN were reported in \cite{Ts}.  This model was reproduced and trained for use as our baseline for the horizon $k=100$.  The baseline training and test curves are shown in Figure~\ref{figure:acc_cnn} of Appendix~\ref{appendix:baseline}.  In \cite{Ts2} a CNN was combined with an LSTM resulting in the architecture denoted CNN-LSTM.  An improvement over the CNN-LSTM architecture, named DeepLOB, was achieved in \cite{Zh} by using an inception module between the CNN and LSTM together with a different choice of convolution filters, stride and pooling.  Finally, the architecture C(TABL) refers to the best performing implementation of the temporal attention augmented bilinear network of \cite{Tr}.  

Our results are shown in Table~\ref{table:horizon10}, Table~\ref{table:horizon20}, Table~\ref{table:horizon50} and Table~\ref{table:horizon100} for each of the horizon choices respectively.  The training and test curves with respect to accuracy for $k=100$ are shown in Figure~\ref{figure:acc_translob} of Appendix~\ref{appendix:baseline}.

\begin{table}[!htb]
\scriptsize
\begin{center}
\begin{tabular}{|p{3cm}||p{2cm}|p{2cm}|p{2cm}|p{2cm}|}
\hline
Model  & Accuracy & Precision & Recall & F1\\
 \hline\hline
SVM \cite{Ts3}  &-  & 39.62 & 44.92 & 35.88 \\
MLP \cite{Ts3}    &-  & 47.81 &  60.78 &  48.27\\
CNN \cite{Ts}  &-  & 50.98 & 65.54 & 55.21 \\
LSTM \cite{Ts3}   & - & 60.77 & 75.92 & 66.33  \\
CNN-LSTM \cite{Ts2} &  - & 56.00 & 45.00 & 44.00 \\
C(TABL) \cite{Tr}  & 84.70 & 76.95 & 78.44 & 77.63 \\
DeepLOB \cite{Zh} &  84.47 & 84.00 &  84.47 & 83.40 \\
TransLOB  & \textbf{87.66} & \textbf{91.81} & \textbf{87.66} & \textbf{88.66} \\
 \hline
\end{tabular}
\end{center}
\caption{Prediction horizon $k=10$.}
\label{table:horizon10}
\end{table}

\begin{table}[!htb]
\scriptsize
\begin{center}
\begin{tabular}{|p{3cm}||p{2cm}|p{2cm}|p{2cm}|p{2cm}|}
\hline
Model  & Accuracy & Precision & Recall & F1\\
 \hline\hline
SVM \cite{Ts3}  & - & 45.08  & 47.77 & 43.20 \\
MLP \cite{Ts3}    &-  & 51.33 &  65.20 & 51.12 \\
CNN \cite{Ts}  & - & 54.79 & 67.38  &  59.17 \\
LSTM \cite{Ts3}   &-  & 59.60 & 70.52 & 62.37 \\
CNN-LSTM \cite{Ts2} & - & - & - & - \\
C(TABL) \cite{Tr}  & 73.74 & 67.18 & 66.94 &  66.93 \\
DeepLOB \cite{Zh} & 74.85 & 74.06 & 74.85 & 72.82 \\
TransLOB  &  \textbf{78.78} & \textbf{86.17} & \textbf{78.78} & \textbf{80.65} \\
 \hline
\end{tabular}
\end{center}
\caption{Prediction horizon $k=20$.}
\label{table:horizon20}
\end{table}

\begin{table}[!htb]
\scriptsize
\begin{center}
\begin{tabular}{|p{3cm}||p{2cm}|p{2cm}|p{2cm}|p{2cm}|}
\hline
Model  & Accuracy & Precision & Recall & F1\\
 \hline\hline
SVM \cite{Ts3}  & - & 46.05 & 60.30 & 49.42 \\
MLP \cite{Ts3}    & - & 55.21 & 67.14 & 55.95 \\
CNN \cite{Ts}  & - & 55.58 & 67.12 & 59.44 \\
LSTM \cite{Ts3}   &-  & 60.03 & 68.58 & 61.43 \\
CNN-LSTM \cite{Ts2} & - & 56.00 & 47.00  &  47.00 \\
C(TABL) \cite{Tr}  & 79.87 & 79.05 & 77.04 &  78.44 \\
DeepLOB \cite{Zh} & 80.51 & 80.38 & 80.51 & 80.35 \\
TransLOB  & \textbf{88.12} & \textbf{88.65} & \textbf{88.12} & \textbf{88.20} \\
 \hline
\end{tabular}
\end{center}
\caption{Prediction horizon $k=50$.}
\label{table:horizon50}
\end{table}

\begin{table}[!htb]
\scriptsize
\begin{center}
\begin{tabular}{|p{3cm}||p{2cm}|p{2cm}|p{2cm}|p{2cm}|}
\hline
Model  & Accuracy & Precision & Recall & F1\\
 \hline\hline
CNN \cite{Ts}  & 63.06 & 63.29 & 63.06 & 62.97 \\
TransLOB  & \textbf{91.62} & \textbf{91.63} & \textbf{91.62} & \textbf{91.61}  \\
 \hline
\end{tabular}
\end{center}
\caption{Prediction horizon $k=100$.}
\label{table:horizon100}
\end{table}

For inspection of our model, we plot the attention distributions for all three heads in the first transformer block.  A random sample input was chosen from the horizon $k=10$ test set.  Pixel intensity has been scaled for ease of visualization.  The vertical axes represent the query index $0\leq i\leq 100$ and the horizontal axes represent the key index $0\leq j\leq 100$.  Queries are aware of the distance to keys through the position embedding layer and entities are only updated with memory from the past owing to the attention mask.  As can be seen in Figure~\ref{figure:headone}, and Figure~\ref{figure:headtwo} and Figure~\ref{figure:headthree} of Appendix~\ref{appendix:attentiondist}, the different heads learn to attend to different properties of the temporal dynamics.  A majority of the queries pay special attention to the most recent keys which is sensible for predicting the next price movement.  This is particularly clear in heads two and three.

\begin{figure}[h]
\begin{center}
\includegraphics[scale=0.3, trim = 0.8cm 2cm 0.8cm 1cm, clip]{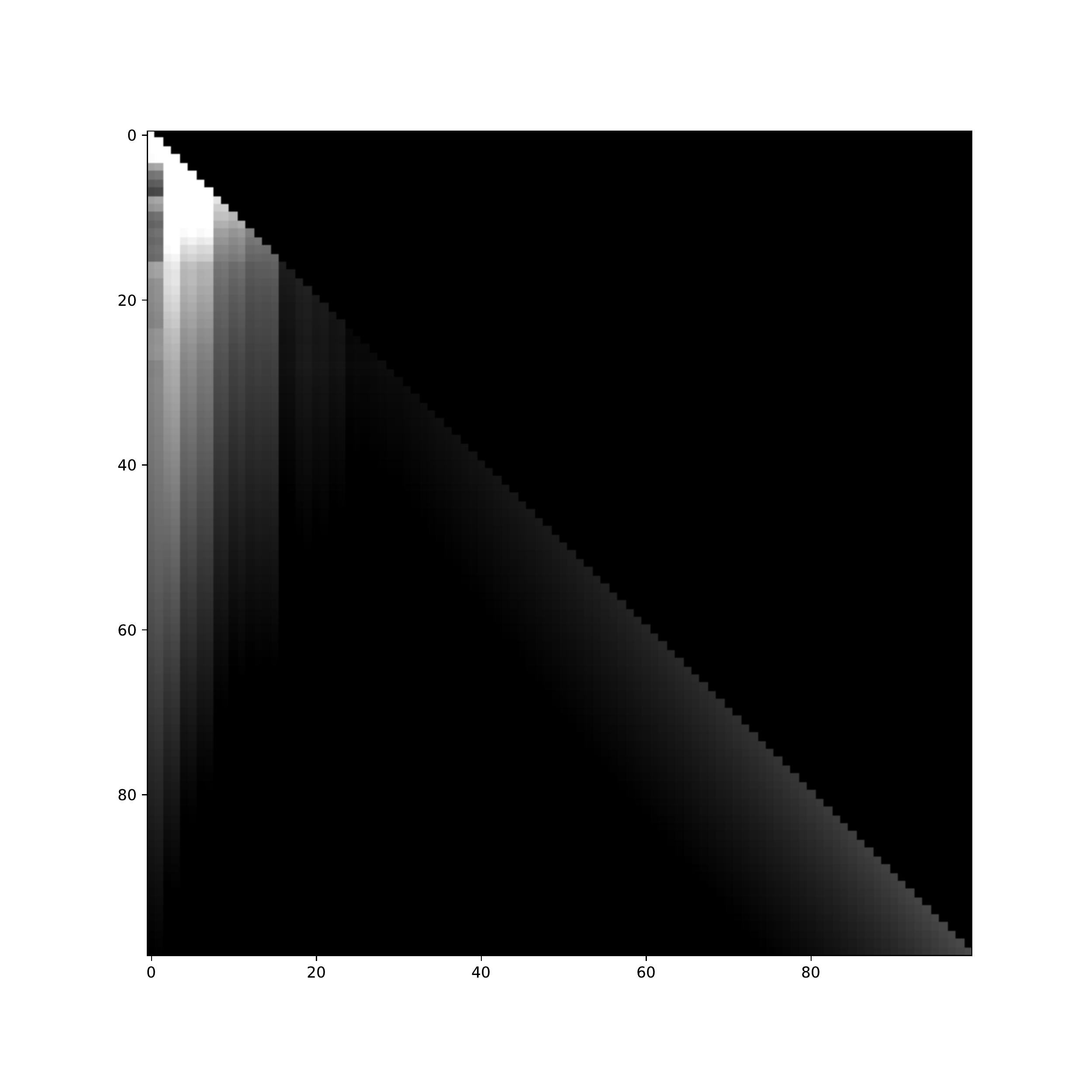}  
\end{center}
\caption{First head of the first transformer block.}
\label{figure:headone}
\end{figure}

\section{Discussion}

We have shown that the limit order book contains informative information to enable price movement prediction using deep neural networks with a causal and relational inductive bias.  This was shown by introducing the architecture TransLOB which contains both a dilated causal convolutional module and a masked transformer module.  This architecture was tested on the publicly available FI-2010 dataset achieving state-of-the-art results.  We expect further improvements using more sophisticated proprietary additions such as the inclusion of sentiment information from news, social media and other sources.  However, this work was developed to exploit only the information contained in the LOB and serves as very strong baseline from which additional tools can be added.  

Due to the limited nature of the FI-2010 dataset, significant time was spend tuning hyperparameters of our model to negate overfitting.  In particular, our architecture was notably sensitive to the initialization.  However, due to the very strong performance of the model, together with the flexibility and sensible inductive biases of the architecture, we expect robust results on larger LOB datasets.  This is an important second step and will be addressed in future work.  In particular, this will allow us to explore the generalization capabilities of the model together with the optimization of important parameters such as the horizon $k$ and threshold $\alpha$.  Nevertheless, based on these initial results we argue that further investigation of transformer based models for financial time-series prediction tasks is warranted.

The efficiency of our algorithm is another imporant property which makes it amenable to training on larger datasets and LOB data with larger event windows.  In spite of the $O(N^2)$ complexity of the self-attention component, our architecture is significantly more sample efficient than existing LSTM architectures for this task such as \cite{Ts3,Ts2,Zh}.  However, moving far beyond the window size of 100, to the territory of LOB datasets on the scale of months or years, it would be interesting to explore sparse and compressed representations in the transformer blocks.  Implementations of sparsity and compression can be found in \cite{Chi,Su,Lam,Li} and \cite{Ki,Rae} respectively.  

Looking forward, similar to recent advances in natural language processing, the next generation of financial time-series models should implement self-supervision as pretraining \cite{De,Ra}.  Finally, it would be interesting to consider the influence of higher-order self-attention \cite{Cl} in LOB and other financial time-series applications.

\section*{Acknowledgements}

The author would like to thank Andrew Royal and Zihao Zhang for correspondence related to this project.


\bibliography{TransLOB} 

\begin{thebibliography}{10}

\bibitem{Ar}
A.~A. Ariyo, A.~O. Adewumi, and C.~K. Ayo.
\newblock Stock price prediction using the arima model.
\newblock In {\em 2014 UKSim-AMSS 16th International Conference on Computer
  Modelling and Simulation}, pages 106--112. IEEE, 2014.

\bibitem{Ba}
J.~L. Ba, J.~R. Kiros, and G.~E. Hinton.
\newblock Layer normalization.
\newblock {\em arXiv preprint arXiv:1607.06450}, 2016.

\bibitem{Bah}
D.~Bahdanau, K.~Cho, and Y.~Bengio.
\newblock Neural machine translation by jointly learning to align and
  translate.
\newblock {\em arXiv preprint arXiv:1409.0473}, 2014.

\bibitem{Bao}
W.~Bao, J.~Yue, and Y.~Rao.
\newblock A deep learning framework for financial time series using stacked
  autoencoders and long-short term memory.
\newblock {\em PloS one}, 12(7), 2017.

\bibitem{Bat}
P.~W. Battaglia, J.~B. Hamrick, V.~Bapst, A.~Sanchez-Gonzalez, V.~Zambaldi,
  M.~Malinowski, A.~Tacchetti, D.~Raposo, A.~Santoro, R.~Faulkner, et~al.
\newblock Relational inductive biases, deep learning, and graph networks.
\newblock {\em arXiv preprint arXiv:1806.01261}, 2018.

\bibitem{Ch}
K.~Chen, J.~Wang, L.-C. Chen, H.~Gao, W.~Xu, and R.~Nevatia.
\newblock Abc-cnn: An attention based convolutional neural network for visual
  question answering.
\newblock {\em arXiv preprint arXiv:1511.05960}, 2015.

\bibitem{Chi}
R.~Child, S.~Gray, A.~Radford, and I.~Sutskever.
\newblock Generating long sequences with sparse transformers.
\newblock {\em arXiv preprint arXiv:1904.10509}, 2019.

\bibitem{Cl}
J.~Clift, D.~Doryn, D.~Murfet, and J.~Wallbridge.
\newblock Logic and the 2-simplicial transformer.
\newblock In {\em Proceedings of the International Conference on Learning
  Representations}, 2020.

\bibitem{Da}
Z.~Dai, Z.~Yang, Y.~Yang, J.~Carbonell, Q.~V. Le, and R.~Salakhutdinov.
\newblock Transformer-xl: Attentive language models beyond a fixed-length
  context.
\newblock {\em arXiv preprint arXiv:1901.02860}, 2019.

\bibitem{De}
J.~Devlin, M.-W. Chang, K.~Lee, and K.~Toutanova.
\newblock Bert: Pre-training of deep bidirectional transformers for language
  understanding.
\newblock {\em arXiv preprint arXiv:1810.04805}, 2018.

\bibitem{Di}
M.~Dixon.
\newblock Sequence classification of the limit order book using recurrent
  neural networks.
\newblock {\em Journal of computational science}, 24:277--286, 2018.

\bibitem{Do}
J.~Doering, M.~Fairbank, and S.~Markose.
\newblock Convolutional neural networks applied to high-frequency market
  microstructure forecasting.
\newblock In {\em 2017 9th Computer Science and Electronic Engineering (CEEC)},
  pages 31--36. IEEE, 2017.

\bibitem{Fi}
T.~Fischer and C.~Krauss.
\newblock Deep learning with long short-term memory networks for financial
  market predictions.
\newblock {\em European Journal of Operational Research}, 270(2):654--669,
  2018.

\bibitem{Go}
M.~D. Gould, M.~A. Porter, S.~Williams, M.~McDonald, D.~J. Fenn, and S.~D.
  Howison.
\newblock Limit order books.
\newblock {\em Quantitative Finance}, 13(11):1709--1742, 2013.

\bibitem{Gr}
K.~Greff, R.~K. Srivastava, J.~Koutn{\'\i}k, B.~R. Steunebrink, and
  J.~Schmidhuber.
\newblock Lstm: A search space odyssey.
\newblock {\em IEEE transactions on neural networks and learning systems},
  28(10):2222--2232, 2016.

\bibitem{Ho}
S.~Hochreiter and J.~Schmidhuber.
\newblock Long short-term memory.
\newblock {\em Neural computation}, 9(8):1735--1780, 1997.

\bibitem{Ki}
N.~Kitaev, {\L}.~Kaiser, and A.~Levskaya.
\newblock Reformer: The efficient transformer.
\newblock In {\em Proceedings of the International Conference on Learning
  Representations}, 2020.

\bibitem{La}
G.~Lai, W.-C. Chang, Y.~Yang, and H.~Liu.
\newblock Modeling long-and short-term temporal patterns with deep neural
  networks.
\newblock In {\em The 41st International ACM SIGIR Conference on Research \&
  Development in Information Retrieval}, pages 95--104, 2018.

\bibitem{Lam}
G.~Lample, A.~Sablayrolles, M.~Ranzato, L.~Denoyer, and H.~J{\'e}gou.
\newblock Large memory layers with product keys.
\newblock In {\em Advances in Neural Information Processing Systems}, pages
  8546--8557, 2019.

\bibitem{Le}
Y.~LeCun, B.~Boser, J.~S. Denker, D.~Henderson, R.~E. Howard, W.~Hubbard, and
  L.~D. Jackel.
\newblock Backpropagation applied to handwritten zip code recognition.
\newblock {\em Neural computation}, 1(4):541--551, 1989.

\bibitem{Li}
S.~Li, X.~Jin, Y.~Xuan, X.~Zhou, W.~Chen, Y.-X. Wang, and X.~Yan.
\newblock Enhancing the locality and breaking the memory bottleneck of
  transformer on time series forecasting.
\newblock In {\em Advances in Neural Information Processing Systems}, pages
  5244--5254, 2019.

\bibitem{Ma}
Y.~M{\"a}kinen, J.~Kanniainen, M.~Gabbouj, and A.~Iosifidis.
\newblock Forecasting jump arrivals in stock prices: new attention-based
  network architecture using limit order book data.
\newblock {\em Quantitative Finance}, 19(12):2033--2050, 2019.

\bibitem{Nt}
A.~Ntakaris, M.~Magris, J.~Kanniainen, M.~Gabbouj, and A.~Iosifidis.
\newblock Benchmark dataset for mid-price forecasting of limit order book data
  with machine learning methods.
\newblock {\em Journal of Forecasting}, 37(8):852--866, 2018.

\bibitem{Oo}
A.~v.~d. Oord, S.~Dieleman, H.~Zen, K.~Simonyan, O.~Vinyals, A.~Graves,
  N.~Kalchbrenner, A.~Senior, and K.~Kavukcuoglu.
\newblock Wavenet: A generative model for raw audio.
\newblock {\em arXiv preprint arXiv:1609.03499}, 2016.

\bibitem{Qi}
Y.~Qin, D.~Song, H.~Chen, W.~Cheng, G.~Jiang, and G.~Cottrell.
\newblock A dual-stage attention-based recurrent neural network for time series
  prediction.
\newblock {\em arXiv preprint arXiv:1704.02971}, 2017.

\bibitem{Ra}
A.~Radford, K.~Narasimhan, T.~Salimans, and I.~Sutskever.
\newblock Improving language understanding with unsupervised learning.
\newblock {\em Technical report, OpenAI}, 2018.

\bibitem{Rae}
J.~W. Rae, A.~Potapenko, S.~M. Jayakumar, and T.~P. Lillicrap.
\newblock Compressive transformers for long-range sequence modelling.
\newblock In {\em Proceedings of the International Conference on Learning
  Representations}, 2020.

\bibitem{Sc}
F.~Scarselli, M.~Gori, A.~C. Tsoi, M.~Hagenbuchner, and G.~Monfardini.
\newblock The graph neural network model.
\newblock {\em IEEE Transactions on Neural Networks}, 20(1):61--80, 2008.

\bibitem{Sh}
S.-Y. Shih, F.-K. Sun, and H.-y. Lee.
\newblock Temporal pattern attention for multivariate time series forecasting.
\newblock {\em Machine Learning}, 108(8-9):1421--1441, 2019.

\bibitem{So}
H.~Song, D.~Rajan, J.~J. Thiagarajan, and A.~Spanias.
\newblock Attend and diagnose: Clinical time series analysis using attention
  models.
\newblock In {\em Thirty-second AAAI conference on artificial intelligence},
  2018.

\bibitem{Su}
S.~Sukhbaatar, E.~Grave, P.~Bojanowski, and A.~Joulin.
\newblock Adaptive attention span in transformers.
\newblock {\em arXiv preprint arXiv:1905.07799}, 2019.

\bibitem{Sz}
C.~Szegedy, W.~Liu, Y.~Jia, P.~Sermanet, S.~Reed, D.~Anguelov, D.~Erhan,
  V.~Vanhoucke, and A.~Rabinovich.
\newblock Going deeper with convolutions.
\newblock In {\em Proceedings of the IEEE conference on computer vision and
  pattern recognition}, pages 1--9, 2015.

\bibitem{Tr}
D.~T. Tran, A.~Iosifidis, J.~Kanniainen, and M.~Gabbouj.
\newblock Temporal attention-augmented bilinear network for financial
  time-series data analysis.
\newblock {\em IEEE transactions on neural networks and learning systems},
  30(5):1407--1418, 2018.

\bibitem{Ts}
A.~Tsantekidis, N.~Passalis, A.~Tefas, J.~Kanniainen, M.~Gabbouj, and
  A.~Iosifidis.
\newblock Forecasting stock prices from the limit order book using
  convolutional neural networks.
\newblock In {\em 2017 IEEE 19th Conference on Business Informatics (CBI)},
  volume~1, pages 7--12. IEEE, 2017.

\bibitem{Ts3}
A.~Tsantekidis, N.~Passalis, A.~Tefas, J.~Kanniainen, M.~Gabbouj, and
  A.~Iosifidis.
\newblock Using deep learning to detect price change indications in financial
  markets.
\newblock In {\em 2017 25th European Signal Processing Conference (EUSIPCO)},
  pages 2511--2515. IEEE, 2017.

\bibitem{Ts2}
A.~Tsantekidis, N.~Passalis, A.~Tefas, J.~Kanniainen, M.~Gabbouj, and
  A.~Iosifidis.
\newblock Using deep learning for price prediction by exploiting stationary
  limit order book features.
\newblock {\em arXiv preprint arXiv:1810.09965}, 2018.

\bibitem{Va}
A.~Vaswani, N.~Shazeer, N.~Parmar, J.~Uszkoreit, L.~Jones, A.~N. Gomez,
  {\L}.~Kaiser, and I.~Polosukhin.
\newblock Attention is all you need.
\newblock In {\em Advances in neural information processing systems}, pages
  5998--6008, 2017.

\bibitem{Vi}
O.~Vinyals, I.~Babuschkin, W.~M. Czarnecki, M.~Mathieu, A.~Dudzik, J.~Chung,
  D.~H. Choi, R.~Powell, T.~Ewalds, P.~Georgiev, et~al.
\newblock Grandmaster level in starcraft ii using multi-agent reinforcement
  learning.
\newblock {\em Nature}, 575(7782):350--354, 2019.

\bibitem{Xu}
K.~Xu, J.~Ba, R.~Kiros, K.~Cho, A.~Courville, R.~Salakhudinov, R.~Zemel, and
  Y.~Bengio.
\newblock Show, attend and tell: Neural image caption generation with visual
  attention.
\newblock In {\em International conference on machine learning}, pages
  2048--2057, 2015.

\bibitem{Za}
V.~Zambaldi, D.~Raposo, A.~Santoro, V.~Bapst, Y.~Li, I.~Babuschkin, K.~Tuyls,
  D.~Reichert, T.~Lillicrap, E.~Lockhart, M.~Shanahan, V.~Langston, R.~Pascanu,
  M.~Botvinick, O.~Vinyals, and P.~Battaglia.
\newblock Deep reinforcement learning with relational inductive biases.
\newblock In {\em Proceedings of the International Conference on Learning
  Representations}, 2019.

\bibitem{Zh}
Z.~Zhang, S.~Zohren, and S.~Roberts.
\newblock Deeplob: Deep convolutional neural networks for limit order books.
\newblock {\em IEEE Transactions on Signal Processing}, 67(11):3001--3012,
  2019.

\bibitem{Zi}
E.~Zivot and J.~Wang.
\newblock Vector autoregressive models for multivariate time series.
\newblock {\em Modeling Financial Time Series with S-Plus}, pages 385--429,
  2006.

\end{thebibliography}
\bibliographystyle{abbrv}

\newpage
\appendix

\section{Training curves}\label{appendix:baseline}

\begin{figure}[!h]
\begin{center}
\includegraphics[scale=0.4]{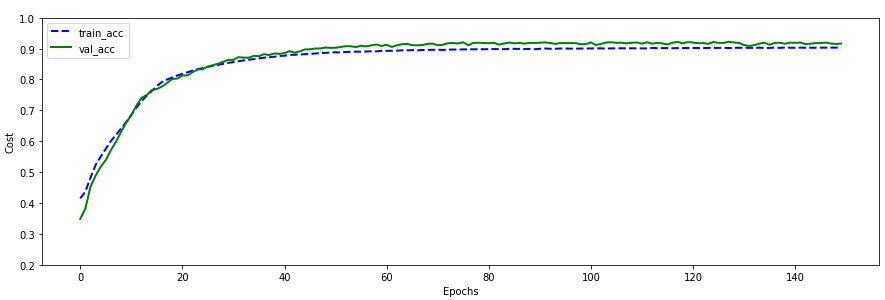}  
\end{center}
\caption{Training and validation accuracy for TransLOB for $k=100$.}
\label{figure:acc_translob}
\end{figure}

\begin{figure}[!h]
\begin{center}
\includegraphics[scale=0.4]{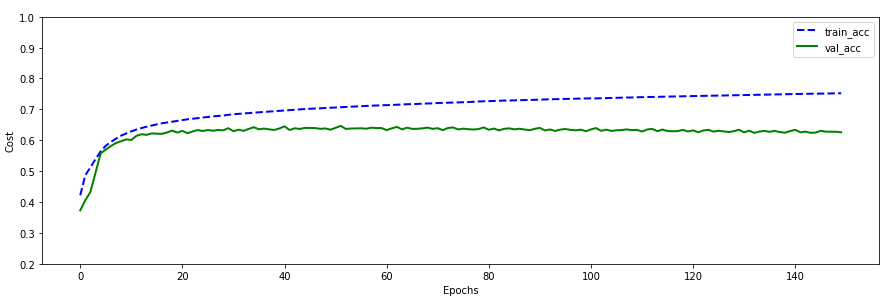}  
\end{center}
\caption{Training and validation accuracy for baseline CNN for $k=100$.}
\label{figure:acc_cnn}
\end{figure}

We plot the training and validation history with respect to accuracy for both our TransLOB architecture in Figure~\ref{figure:acc_translob} and the baseline CNN architecture of \cite{Ts} in Figure~\ref{figure:acc_cnn}.

\section{Attention distributions}\label{appendix:attentiondist}

We include here the remaining visualizations of the attention output of our learned model in the first transformer block.  Input is a random sample for the horizon $k=10$.  

\begin{figure}[!htb]
\begin{center}
\includegraphics[scale=0.3, trim = 0.8cm 2cm 0.8cm 1cm, clip]{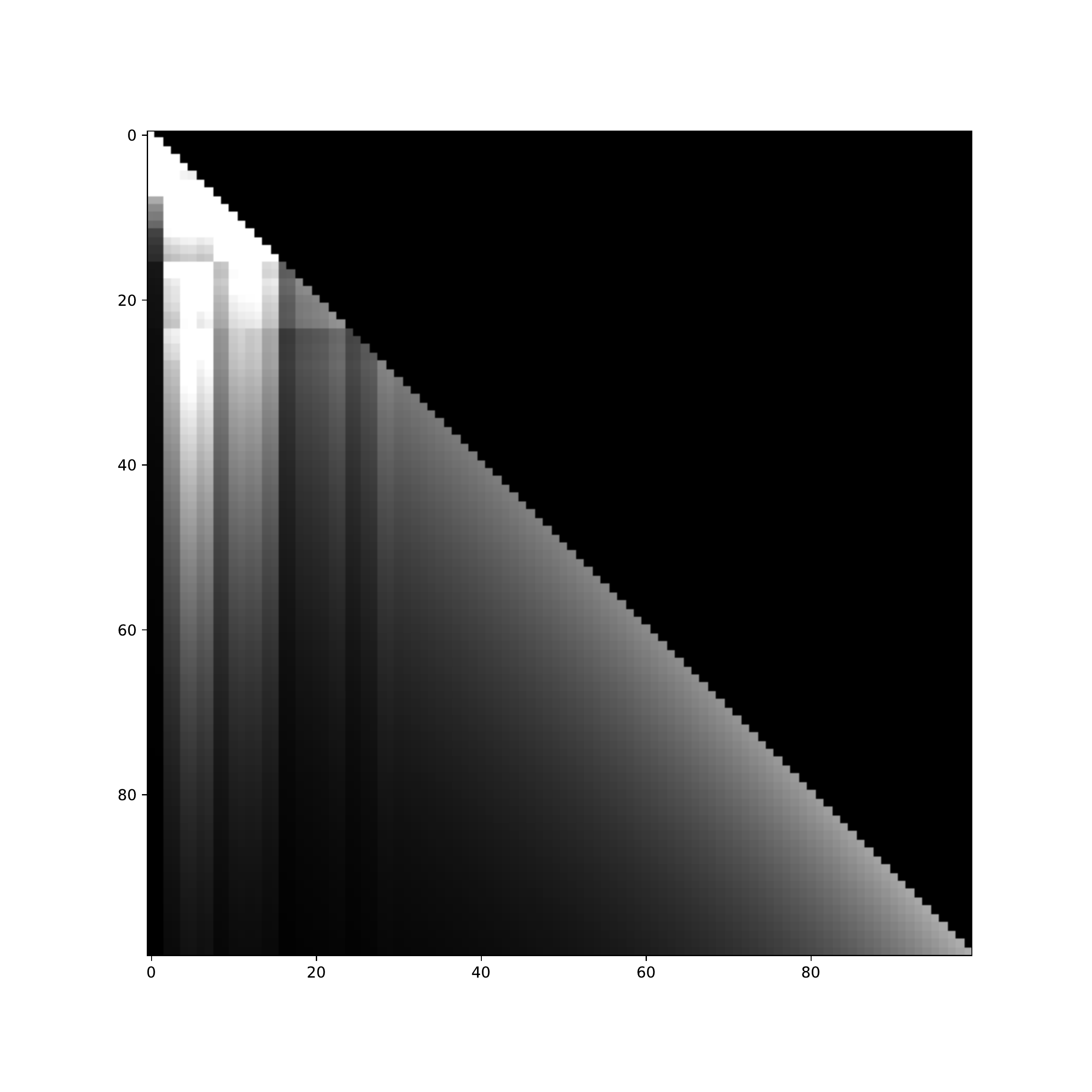}  
\end{center}
\caption{Second head of the first transformer block.}
\label{figure:headtwo}
\end{figure}

\begin{figure}[!htb]
\begin{center}
\includegraphics[scale=0.3, trim = 0.8cm 2cm 0.8cm 1cm, clip]{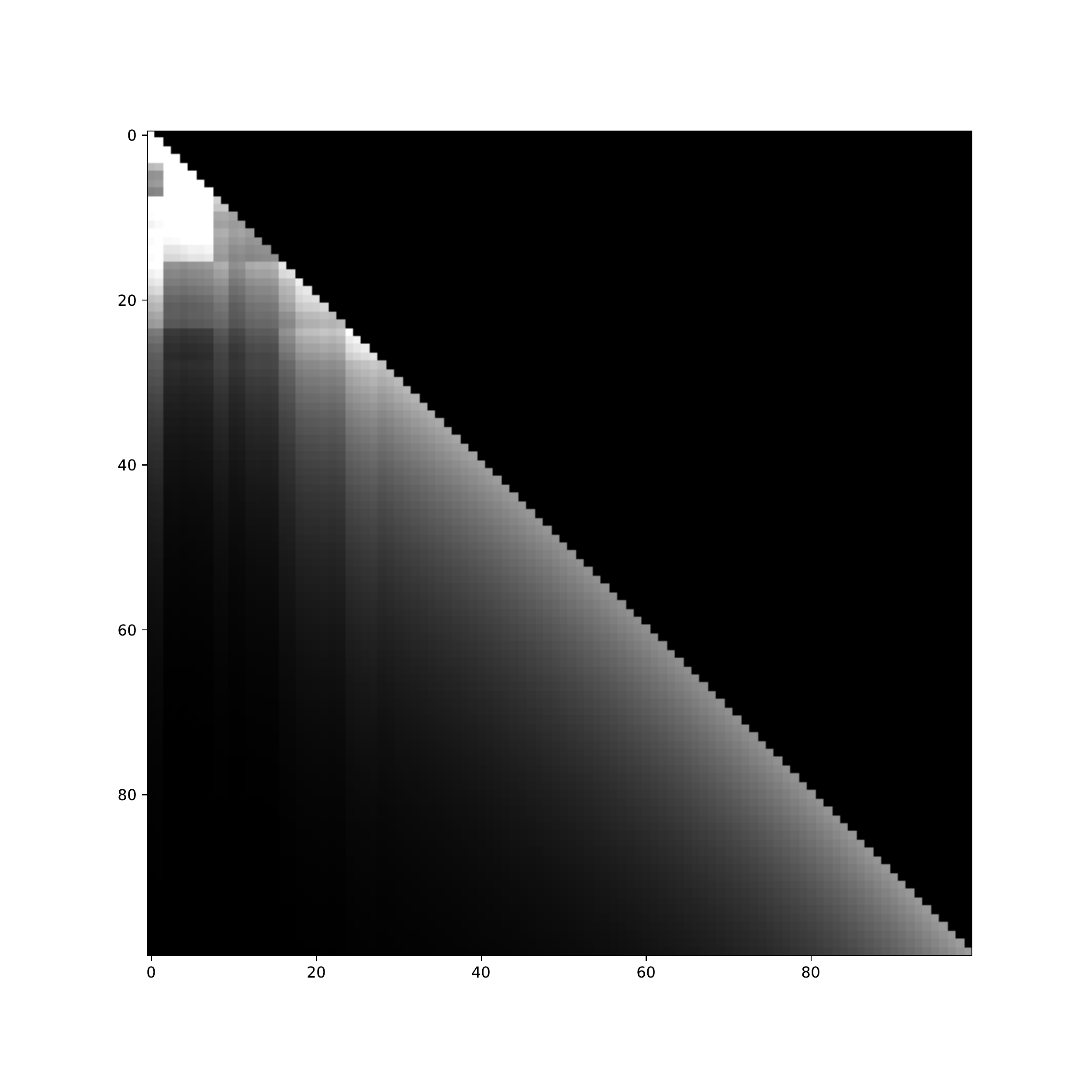}  
\end{center}
\caption{Third head of the first transformer block.}
\label{figure:headthree}
\end{figure}

\end{document}